# Overview of quantitative susceptibility mapping using deep learning - Current status, challenges and opportunities

## Woojin Jung[a*], Steffen Bollmann[b,c*], Jongho Lee[a]

[a]Laboratory for Imaging Science and Technology, Department of Electrical and Computer Engineering, Seoul National University, Seoul, South Korea
[b]ARC Training Centre for Innovation in Biomedical Imaging Technology, The University of Queensland, Building 57 of University Dr, St Lucia QLD 4072, Brisbane, Australia
[c]Centre for Advanced Imaging, The University of Queensland, Building 57 of University Dr, St Lucia QLD 4072, Brisbane, Australia

**\*Shared first authors**

**Corresponding authors:**
Jongho Lee (jonghoyi@snu.ac.kr) & Steffen Bollmann (steffen.bollmann@cai.uq.edu.au)

## Competing Interests statement

Author SB is co-inventor of a patent "Solving the ill-posed quantitative susceptibility mapping inverse problem using deep convolutional neural networks" (US 2019 / 0204401 A1)

## Acknowledgements

The authors acknowledge the facilities and scientific and technical assistance of the National Imaging Facility, a National Collaborative Research Infrastructure Strategy (NCRIS) capability, at the Centre for Advanced Imaging, The University of Queensland. This work was also supported by the National Research Foundation of Korea grant (NRF-2018R1A2B3008445) and Institute of Engineering Research and Interdisciplinary Research Initiatives Program by College of Engineering and College of Medicine, Seoul National University.

## Word count

3202

# Abstract

Quantitative susceptibility mapping (QSM) has gained broad interests in the field by extracting biological tissue properties, predominantly myelin, iron and calcium from magnetic resonance imaging (MRI) phase measurements in vivo. Thereby, QSM can reveal pathological changes of these key components in a variety of diseases. QSM requires multiple processing steps such as phase unwrapping, background field removal and field-to-source-inversion. Current state of the art techniques utilize iterative optimization procedures to solve the inversion and background field correction, which are computationally expensive and require a careful choice of regularization parameters. With the recent success of deep learning using convolutional neural networks for solving ill-posed reconstruction problems, the QSM community also adapted these techniques and demonstrated that the QSM processing steps can be solved by efficient feed forward multiplications not requiring iterative optimization nor the choice of regularization parameters. Here, we review the current status of deep learning based approaches for processing QSM, highlighting limitations and potential pitfalls, and discuss the future directions the field may take to exploit the latest advances in deep learning for QSM.

# Introduction

Quantitative susceptibility mapping (QSM) aims to compute the spatial distribution of magnetic susceptibility from the signal phase of gradient echo magnetic resonance imaging (MRI) data.[1,2] Magnetic susceptibility describes the degree of magnetization of a material in a magnetic field and QSM thereby delivers non-invasive insights into tissue composition and microstructure.[3,4] In the brain, QSM has shown to provide information about myelin[5–7] and iron concentration.[8,9] Most applications have been in brain imaging to study normal aging,[10] Huntington's Disease,[11] Multiple Sclerosis,[12] Alzheimer's Disease[13,14] and Parkinson's Disease[15] and to allow visualization and differentiation of blood depositions from calcifications.[16] Applications outside the brain include quantifying liver iron[17,18] and imaging cartilage in joints.[19]

Obtaining a quantitative susceptibility map requires a gradient-recalled-echo sequence[20] or gradient-echo-based echo-planar imaging[21,22] where the signal phase is related to local magnetic field changes. In order to compute the magnetic susceptibility, the raw signal phase is first unwrapped, and magnetic field changes from regions outside the object of interest are removed before the field perturbation can be related to the underlying tissue magnetic susceptibility distribution by solving an ill-posed inverse problem.[23]

Recently, deep learning or deep convolutional neural networks have been demonstrated as a powerful tool to solve image processing problems, including inverse problems[24] and MRI reconstruction.[25–27] The use of neural networks in solving inverse problems is motivated by the fact that a neural network can approximate any continuous function when the network has enough free parameters.[28] Furthermore, the network automatically learns necessary features for data processing and, therefore, does not require explicit feature selection that may be sub-optimal. As a result, deep convolutional neural networks have outperformed human-designed and classical machine learning algorithms in many image reconstruction problems.[29–31] An additional practical advantage of neural networks is the computational efficiency when generating a feed forward output.[24] In particular, the use of graphic processing units has substantially increased the computational efficiency. All of these advantages may provide potential improvements in QSM reconstruction.

The aim of this work is to review the current deep learning approaches for processing QSM, highlighting limitations and potential pitfalls, and to discuss the possible directions the field may take to exploit the latest advances in deep learning for QSM.

# QSM Processing Pipeline

QSM processing includes multiple steps such as phase unwrapping, background field removal and dipole inversion (Figure 1). The background field removal requires a mask separating the organ of interest from the background and therefore a masking step is necessary in most QSM pipelines. In the following, we will summarize the steps in the QSM pipeline and illustrate where deep learning techniques (summarized in Table 1) can offer new opportunities. To distinguish the different training datasets currently utilized in the literature for training, we define "in-vivo data" to be acquired data from an MRI scanner, where the input is the measured field and the target is a QSM reconstruction using techniques such as COSMOS or traditional optimization techniques. "Simulated in-vivo data" or simply "simulated data" refers to training data that was generated from in-vivo data by computing the forward dipole model to yield the input field. "Synthetic data" refers to training that does not utilize any in-vivo data and is purely trained based on computer-simulated geometrical shapes.

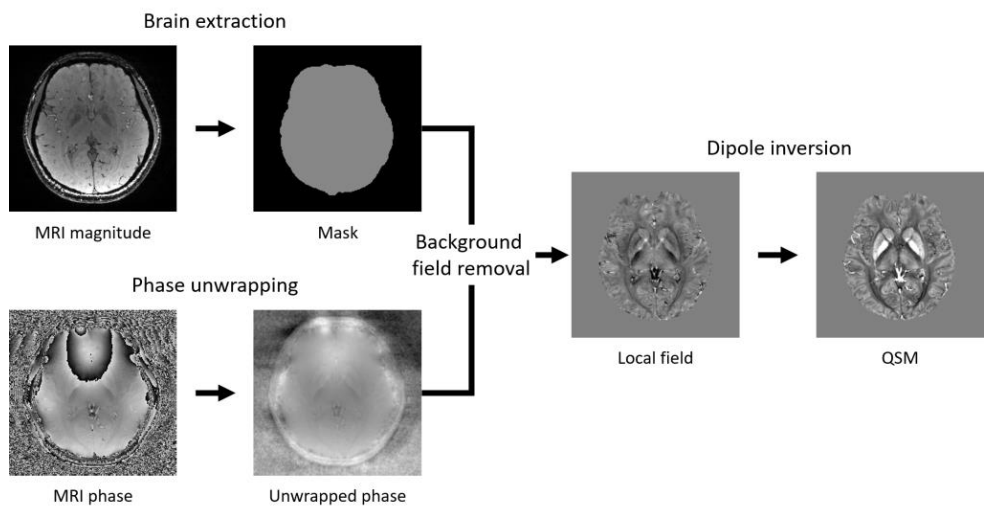

**Figure 1** - The QSM processing pipeline includes phase unwrapping, background field removal (requiring a mask separating background and organ of interest) and the dipole inversion.

## Phase Unwrapping

Phase unwrapping strategies are necessary because MRI detects signal phase between -π and π, leading to phase jumps of 2π in the data. These phase jumps can be corrected by imposing temporal or spatial constraints. An overview of these techniques can be found in the review by Robinson et al.[32] None of these traditional techniques is perfect and selecting a method results in a trade-off between speed and accuracy. Deep learning has the potential to drastically speed up unwrapping and deliver an accurate result. In the presence of rapid phase fluctuations and noise, however, it can be computationally expensive and fragile. Furthermore, application of deep learning for phase unwrapping is not straight forward, because standard network architectures and loss functions do not work well, because phase wraps are not locally constrained to a neighbourhood as assumed by convolutional neural nets with a fixed receptive field size.[33] The idea of applying deep learning for phase unwrapping was first introduced for flow and phase contrast imaging.[34,35] For phase imaging, Ryu et al.[33] developed a novel network with bidirectional recurrent neural network (RNN) modules to learn global features and designed a loss function based on the contrast of the error. The results of this network showed that it can handle global phase wraps and successfully generated unwrapped phase images in a few seconds of time.

## Background Field Removal

The background field in QSM is induced by magnetic field gradients outside the object of interest, e.g. due to tissue-air interfaces, or $B_0$ inhomogeneities due to imperfect shimming.[36] These external fields are orders of magnitude stronger and overlap with the local tissue field changes of interest.

A comprehensive review of traditional background field correction techniques[37] classified the assumptions in 1) methods assuming no sources close to boundaries, 2) methods assuming no harmonic internal and boundary fields, and 3) methods not employing an explicit boundary assumption, but minimizing an objective function. One example assuming no sources close to boundaries is sophisticated harmonic artifact reduction for phase data (SHARP),[38] which was later extended to V-SHARP[39] by reducing artifacts at the edges using decreasing kernel sizes towards the boundary. Another method solves the Laplacian boundary value problem (LBV),[40] assuming no harmonic internal and boundary fields. The boundary condition assumption works in most cases, but can be problematic when the local field varies rapidly near the boundary (e.g. veins close to the brain surface). One example in the third group is regularization enabled SHARP (RESHARP),[41] which utilizes a Tikhonov regularization. Methods that are based on physical properties of dipole sources outside the object of interest, such as projection onto dipole fields (PDF),[36,42] also fall in this category.

The traditional background field corrections perform well given that the regularization parameters are carefully adjusted and assumptions of the boundary conditions are not violated. Most methods face the limitation of a loss of information at the boundaries and they require the definition of a mask, separating the object of interest from the background. However, this mask generation is non-trivial (especially in the abdomen or heart) and leads to the loss of information close to boundary regions or residual artifacts due to a violation of the boundary condition. The need for carefully choosing regularization parameters and defining a region of interest currently limit the robustness and wide clinical applications of QSM.

Deep learning based techniques can offer an increase in robustness and speed-up in computation. For example, SHARQnet[43] was designed for background field removal using a 3D convolutional neural network and was trained on synthetic background fields overlaid on top of a brain simulation. When the performance was compared to SHARP, RESHARP and V-SHARP, SHARQnet delivered accurate background field corrections on simulations and on in-vivo data. In particular, SHARQnet does not need a brain mask defining the object of interest, providing the benefit of omitting the error-prone brain masking step from the QSM pipeline. This would provide advantages in clinical applications as has already been shown in QSM algorithms that can invert the total field.[44,45] A study by Liu et al.[46] also proposed background field removal with similar results using simulated background field distributions.

**Dipole Inversion**

The ill-posed dipole inversion step is traditionally overcome either by additional measurements or by numerical stabilization strategies. Utilizing the acquisition of different orientations with respect to the static magnetic field is known as Calculation of susceptibility through multiple orientation sampling (COSMOS)[47] and requires at least three different orientations to make the field-to-susceptibility problem over-determined and enables an analytical solution. Although COSMOS generates susceptibility maps of high fidelity, and is therefore considered a gold standard for QSM, it assumes isotropic magnetic susceptibility and contains little information about anisotropic tissue properties.[5,7,48] Methods such as susceptibility tensor imaging (STI)[49] or the Generalized Lorentzian Tensor Approach (GLTA)[50] have been developed that extend the magnetic susceptibility scalar to a tensor. Common to all multi-orientation methods is the clinical in-feasibility due to patient discomfort and scan time requirements.

To overcome the limitations, methods have been developed that compute magnetic susceptibility from single orientation data by employing numerical stabilization techniques.

Numerical strategies can be subdivided into inverse filtering and iterative methods.[37] Inverse filtering solves the problem in Fourier domain by dividing the pre-processed phase data by the unit dipole response yielding the magnetic susceptibility. However, small values in the unit dipole response result in an amplification of noise and errors, necessitating the replacement of small values by a fixed threshold. This method is known as truncated k-space division (TKD)[2] and the results can be corrected for underestimated magnetic susceptibility.[37]

Alternatively, the inverse problem can be solved in the spatial domain by reformulating the problem as a Bayesian reconstruction with a data consistency term of the forward dipole model (i.e. convolution of the dipole kernel with the susceptibility distribution) and a carefully designed regularization term, solving a least-squares problem. One example is the Morphology enabled dipole inversion (MEDI) algorithm which utilizes edge information from magnitude images as the regularization term.[51] Another one is the LSQR algorithm[52] in STI Suite where streaking artifacts are first estimated and then subtracted from the initial susceptibility solution (iLSQR[53] and STAR-QSM[54]). Many other QSM algorithms are based on the Bayseian reconstruction and differ in the regularization term that incorporates prior information about the susceptibility distribution.

Deep learning methods for the dipole inversion offer the advantage that they introduce data-driven and self-regulated reconstruction when a deep neural network, typically a 3D Unet (Table 1), is end-to-end trained using local field and QSM pairs. A few networks have successfully demonstrated potentials of learning dipole inversion, generating high quality QSM maps with well-suppressed streaking artifacts (Figure 2). Examples of such networks are QSMnet[55] and QSMnet[+],[56] which learned the relationship between local field and susceptibility maps using in-vivo experimental datasets and simulated in-vivo datasets via end-to-end training of a 3D Unet. For these networks, the gold-standard COSMOS maps were used as the training data, intending to generate COSMOS-quality QSM maps as the output for a single orientation local field input (Figure 2). In addition, the loss function included a forward dipole model term to enforce the network learning the dipole model. Instead of training in-vivo data, one can train on purely synthetic data as demonstrated in DeepQSM.[57] In this work, a network was trained using pairs of a synthetic 3D susceptibility map of geometric shapes and a field map generated by the dipole forward model. The idea of utilizing synthetic data was further extended by generating susceptibility distributions that mimic the spatial frequency of in-vivo brain[58] and by combining synthetic data and simulated data.[59] Other studies suggested to use variational networks[60,61] and generative adversarial networks[62] for dipole inversion.

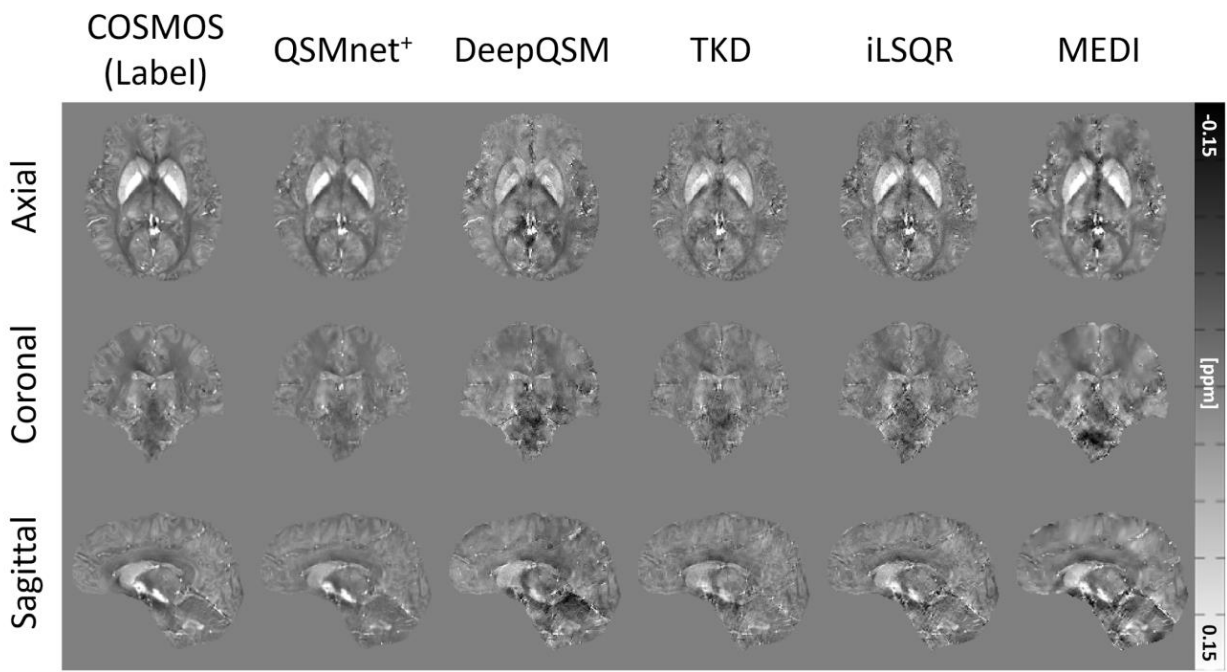

|  | QSMnet+ | DeepQSM | TKD | iLSQR | MEDI |
|---|---|---|---|---|---|
| NRMSE (%) | 51.3 | 101.1 | 69.9 | 78.0 | 96.0 |
| HFEN (%) | 49.5 | 85.5 | 66.0 | 65.3 | 89.0 |
| SSIM | 0.91 | 0.79 | 0.82 | 0.86 | 0.91 |
| pSNR (dB) | 43.3 | 37.4 | 40.6 | 39.6 | 37.8 |

**Figure 2** - A comparison of dipole inversion techniques: From left, COSMOS QSM map (gold standard QSM map using multiple orientation data), two different deep learning-based QSM maps (QSMnet+ trained using in-vivo and simulated data, and DeepQSM trained using synthetic data; single orientation field map is used as input for both methods), and three different conventional QSM maps (TKD, iLSQR, and MEDI; single orientation field map is used as input) are shown. Deep learning QSM maps show superior image quality with little streaking artifacts. Computational times for deep learning methods are a few seconds whereas those of the iterative reconstruction methods (iLSQR and MEDI) are several minutes.

## Phase to QSM

The dipole inversion and background field correction can be solved in a single step with the advantage of higher computation speeds and avoiding error propagation between different optimization procedures. A few conventional methods have already been suggested for this phase to QSM processing.[21,63,64]

Methods solving the background field correction and dipole inversion problem jointly are also promising in deep learning QSM. One approach is to combine separately trained networks, e.g. background field removal and dipole inversion, which has been demonstrated to work in the SHARQnet paper[43] where DeepQSM[57] was used to invert the background field corrected data of SHARQnet. This concept was also suggested in the works by Heber et al.[65] and Kim et al..[66]

Alternatively, a few studies have proposed to directly learn the reconstruction of QSM from phase images in an end-to-end manner. One group proposed deep learning total field inversion where simulated total fields and brain masks were used as the inputs and QSM as the target for training.[59,67] Another approach utilized the unwrapped phase image with no brain mask as the input.[68] These deep learning QSM methods might prevent the potential error propagations from background field removal and dipole inversion by a single-step reconstruction.

# Current Challenges of Deep Learning QSM

In deep learning, one of the biggest issues is generalization errors that happen when test data have different characteristics to training data.[69–71] For example, differences in SNR between test and training data resulted in substantial degradation in image quality when reconstructing undersampled data in MRI.[72] In QSM, such differences can exist in susceptibility range, type of training data (e.g. in-vivo vs. simulated vs. synthetic data), resolution, spatial frequency of image content, SNR, $B_0$ direction and many more. So far, a few studies have demonstrated generalization errors in deep learning QSM. One example is a recent study by Jung et al., reporting the effects of susceptibility range.[56] In the study, the network was trained using healthy volunteer data, limiting the training range to healthy tissue values (Figure 3a). When the network generated the susceptibility map of a hemorrhagic patient, it produced underestimated susceptibility values in hemorrhagic lesions, which have far higher susceptibility values than those of normal tissue (Figure 3). This result demonstrates the importance of the trained susceptibility range in deep learning QSM to achieve generalization capabilities for pathological conditions.

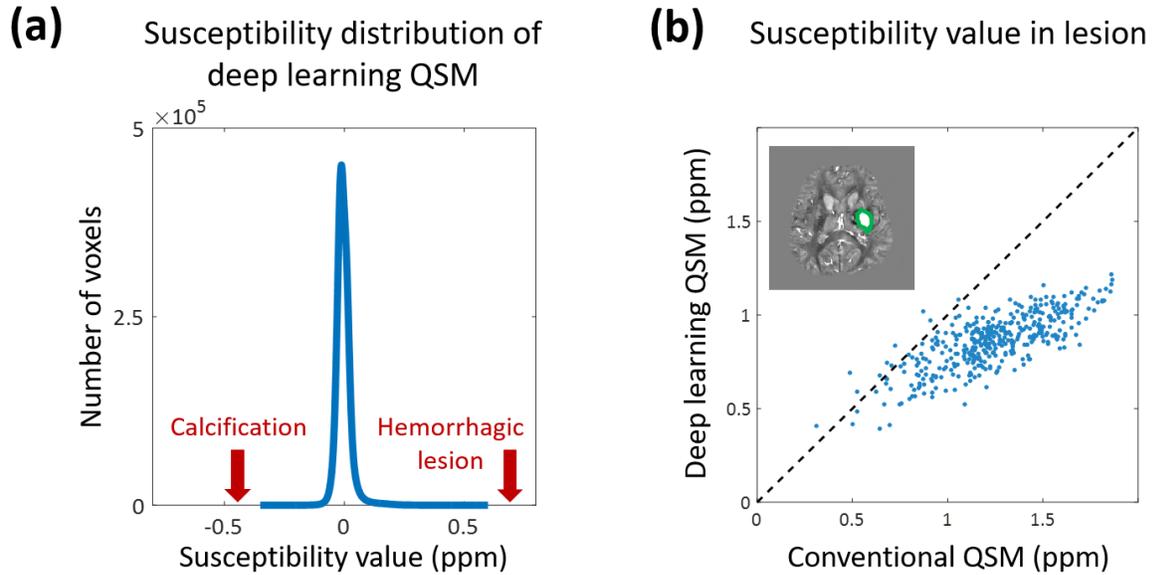

**Figure 3** - The effects of the susceptibility range in deep learning QSM. (a) The susceptibility distribution of healthy volunteers used for network training. Since only healthy volunteer data are used for training, the susceptibility values of pathological tissues (e.g. hemorrhage or calcification) are not covered (red arrows). (b) The scatter plot of the susceptibility values in the hemorrhagic lesion (green circle) reconstructed by conventional QSM (x-axis) and deep learning QSM (QSMnet; y-axis). Compared to the conventional QSM, deep learning QSM results underestimated the susceptibility values of the lesion.

Another consideration for generalization is the type of the training data (e.g. in-vivo vs. simulated data). Many deep learning QSM networks are trained either by in-vivo data or simulated data generated by the dipole convolution of in-vivo susceptibility maps. To demonstrate these effects, we trained three different networks, $\text{Unet}_{\text{In-vivo}}$, $\text{Unet}_{\text{Simul}}$, and $\text{Unet}_{\text{In-vivo+Simul}}$, which were trained on in-vivo data, simulated data, and both in-vivo and simulated data, respectively. The results demonstrate that the performance of the network degrades when the test data deviate from the training data (Figure 4). Another observation is that the network trained using both types of training data shows good performance for both types of test data, suggesting an approach for improving generalization (Figure 4; yellow boxes).

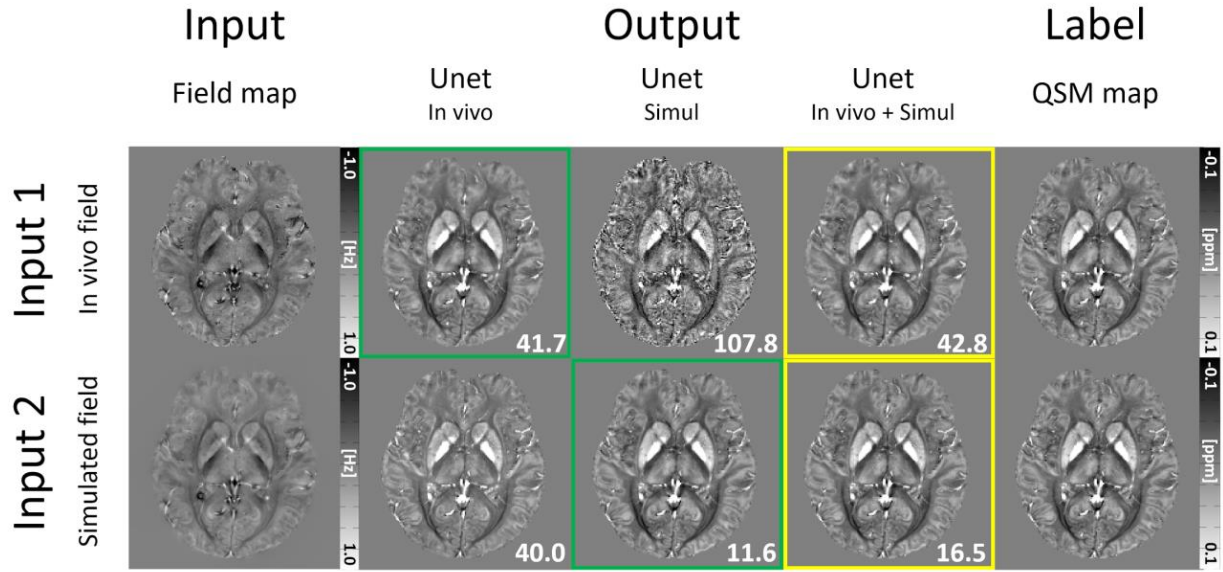

**Figure 4 -** The QSM results of the three networks (Unet$_{In\text{-}vivo}$, Unet$_{Simul}$, and Unet$_{In\text{-}vivo+Simul}$) with different types of training data (in-vivo vs. simulated vs. combined data) are displayed. The simulated field map is generated by the dipole convolution of the COSMOS QSM map. The last column shows the gold standard susceptibility map reconstructed by COSMOS. Normalized root mean squared errors (NRMSE) with respect to the gold standard are displayed in the lower right side of each map. When the type of the input field matches to that of the training data, the networks generated lowest NRMSE results (green boxes). By using both in-vivo and simulated data for training, the network generalizes better for multiple input types (yellow boxes).

Image resolution also plays an important role in the generalization of deep learning QSM. When a network is trained with a fixed resolution, it results in increased errors for data with different resolutions as demonstrated in Figure 5. The performance degradation is particularly severe when the test data have an anisotropic resolution (last column in Figure 5). Hence, a preprocessing step that resamples the input data or a modification of the training data is necessary to achieve optimal results.

When a deep neural network is trained with synthetic data, it is important to generate training data that have enough variability and structure.[73] Additionally, a study suggested that matching the spatial frequency of the synthetic images to that of in-vivo data is important.[74]

Lastly, Høy et al. demonstrated that the performance of deep learning QSM decreases when different noise levels or different $B_0$ directions are applied (Figure 6).[75] The results suggest that matching the orientation of the input data to that of the training data is crucial. For the influence of SNR, one may improve the results by adding simulated noisy data for training.[73]

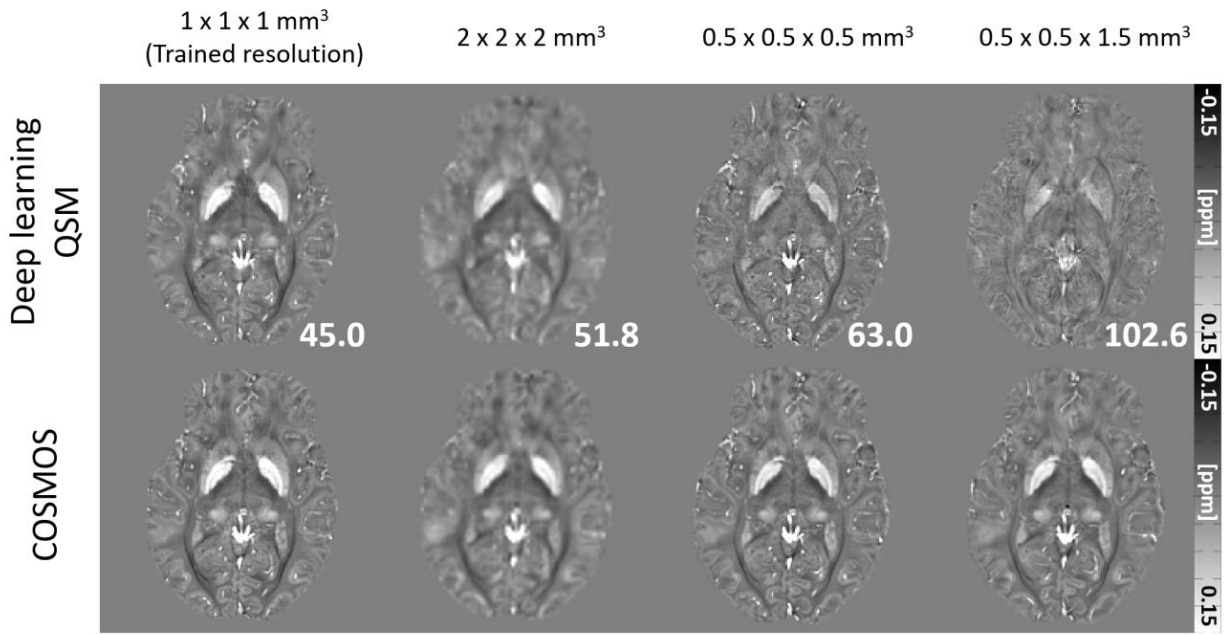

**Figure 5 -** The effects of image resolution on deep learning QSM. The original test data were acquired in 1 mm isotropic resolution and then interpolated into three different resolutions (2 mm isotropic, 0.5 mm isotropic, and 0.5 x 0.5 x 1.5 mm³ resolutions). The performance of the network degrades when the resolution of the test data differs from the training data. NRMSE with respect to the COSMOS map is shown at the bottom left corner of each map.

In summary, the issues listed here for the generalization of deep learning QSM suggest that further investigation is needed to identify the sources of generalization errors. Also, further developments in methods are necessary to improve the generalization, which is the topic of the next chapter of this review.

## (a) Noise level

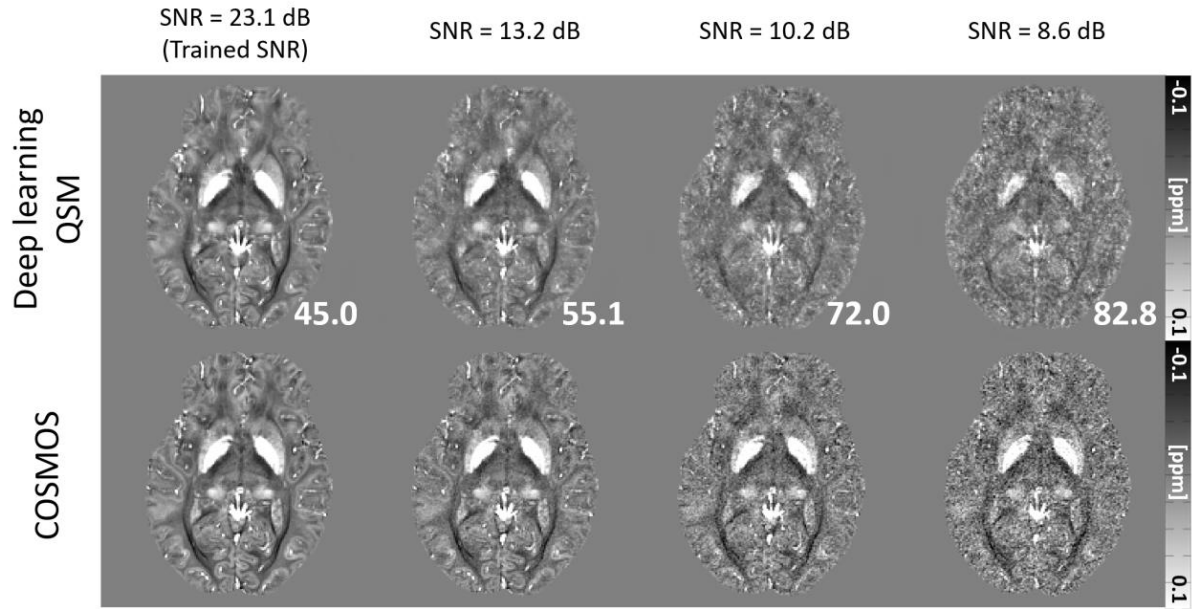

## (b) B₀ direction

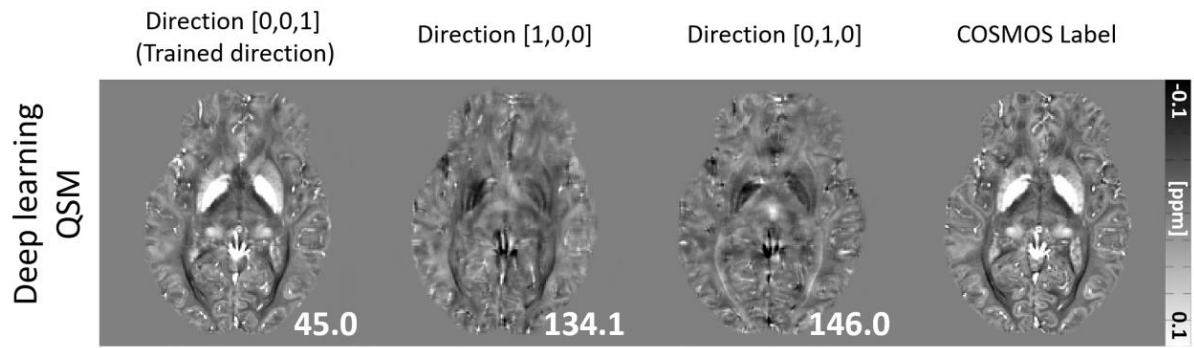

**Figure 6 -** The effects of SNR (a) and $B_0$ direction (b) on deep learning QSM. When the SNR decreases or the $B_0$ direction of test data are different from those of the training data, the performance of the network degrades. NRMSE with respect to the COSMOS map is shown at the bottom left corner of each map.

# Generalization in Deep Learning QSM

To improve generalization in deep learning QSM, a few methods have been proposed. The first approach is to use data augmentation as demonstrated in Figure 4 for the different types of training data. As explained before, the network trained with both in-vivo and simulated data generated better susceptibility maps for both input data types than those trained on a single data type, demonstrating that the network can be generalized for data types using data augmentation. Similarly, Jung et al. showed that data augmentation using susceptibility value scaled maps can cover a wider range of susceptibility values, successfully reconstructing QSM maps for pathology.[56]

Different from other fields, which require additional data acquisition for data augmentation, QSM can be benefited from the dipole model in generating augmentation data. However, as mentioned earlier, the dipole model does not fully reflect in-vivo data and, therefore, one has to be cautious when supplementing model-based data for the training of the network.

The generalization capabilities of deep learning QSM can also be achieved by improving the deep neural network architecture. For example, the variational network concept, which utilizes the network as regularization in the Bayesian reconstruction, has been applied for QSM reconstruction.[60,61] This concept may have potential to achieve better generalization by incorporating deep learning priors in standard gradient descent schemes ensuring data consistency. Similarly, another study utilizes deep learning QSM as the regularization term in the Bayesian reconstruction to improve generalization.[76] Zhang et al. suggested an approach that modifies the weights of a pretrained network using a physics based loss for new datasets, reporting improvement in generalization.[77] Recent studies also suggested to utilize new network architectures with the goal to improve deep learning QSM.[78–80]

# Future Opportunities for Deep Learning QSM

Most of the current implementations of deep learning QSM utilize a simple dipole model with the assumption that magnetic susceptibility is a scalar quantity (i.e. isotropic susceptibility). Deep learning may help to utilize more complex forward models, incorporating additional model terms that account for anisotropy of magnetic susceptibility and microstructure, and chemical exchange. For example, the information of fiber orientation from diffusion measurements may be included as the input of a network to generate susceptibility and microstructural anisotropy maps. Recently, a new model that is not limited by the Lorentzian sphere approximation was proposed (QUASAR; Schweser and Zivadinov, 2018). This model finds frequency contributions unrelated to the spatial variation of bulk magnetic susceptibility in addition to the contributions that adhere to the Lorentzian sphere model and a deep learning approach has shown promising results in separates the two (DEEPOLE; Jochmann et al., 2019b).

Applications of deep learning QSM outside of the brain such as abdomen and knee may open new opportunities for clinical applications. Currently, however, only a few conventional techniques have been proposed. The new deep learning techniques that do not require a mask for the organ of interest could be helpful. Additionally, the network may be trained to separate water and fat for QSM reconstruction in the areas that contain both water and fat.[18]

The deep learning field profited significantly from sharing models, codes and data and the QSM field is following this example. In order to foster sharing of tools and data, we created a GitHub organization named dlQSM that welcomes everyone to contribute their implementations: https://github.com/dlQSM/. This effort will be particularly important when evaluating a new method with existing deep learning networks since the network performance is dependent not only on the network structure but also on the training data. Hence, sharing a trained network along with training data has critical importance.

## Conclusions

Deep learning offers a new angle to QSM reconstruction and has already shown promising results to overcome a few important problems in conventional methods. However, new challenges that are not yet resolved have risen, opening novel opportunities not only for QSM reconstruction but also for deep learning in general. We believe substantial amount of work is left ahead to bring the techniques towards robust clinical applications.

**Table 1** - An overview of the current Deep Learning QSM publications listing the training data, the patch size, the training data size, the network architecture and the loss function utilized.

| Article | Training dataset | Details |
|---|---|---|
| **Phase unwrapping** | | |
| Johnson et al.[34] | Simulated phase<br>3D patch with 11 x 11 x 11<br>Training data size: 250,000 | 3D network similar to VGG<br>Loss function: Not reported |
| He et al.[35] | Simulated phase<br>3D patch with 20 x 30 x 30<br>Training data size: Not reported | 3D ResNet<br>Loss function: Not reported |
| Ryu et al.[33] | In vivo and synthetic data<br>2D slice<br>Training data size: not reported | 2D Bidirectional RNN<br>Loss function: Total variation + variation of error |
| **Background Field Removal** | | |
| Bollmann et al.[43] | Synthetic data<br>3D patch with 32 x 32 x 32<br>Training data size: 100,000 | 3D U-net<br>Loss function: L2 loss |
| Liu, J. et al.[46] | Simulated background field<br>3D patch with 256 x 256 x 64<br>Training data size: 6,000 | 3D gated CNN,<br>Loss function: L1 loss + gradient different loss |
| **Dipole inversion** | | |
| Gong et al.[81] | In vivo COSMOS<br>2D multi-contrast inputs<br>data size: 9,600 | 2D U-net<br>Loss function: L1 loss |
| Yoon et al.[55] | In vivo COSMOS and simulated data<br>3D patch with 64 x 64 x 64<br>Training data size: 16,800 | 3D U-net<br>Loss function: L1 loss + gradient different loss<br>+ model loss |
| Bollmann et al.[57] | Synthetic data<br>3D patch with 64 x 64 x 64<br>Training data size: 100,000 | 3D U-net<br>Loss function: L2 loss |
| Chen et al.[62] | In vivo COSMOS<br>3D patch with 64 x 64 x 64<br>Training data size: Not reported | 3D U-net based Generative adversarial network<br>Loss function: L1 loss + content loss + adversarial loss |
| Gao et al.[79] | In vivo QSM and synthetic data<br>3D patch with 48 x 48 x 48<br>Training data size: 15,000 | 3D U-net with octave convolution<br>Loss function: L1 loss |
| Kames et al.[60] | In vivo COSMOS<br>3D patch<br>Training data size: Not reported | 3D Variational network<br>Loss function: Not reported |
| Liu and Koch[82] | Simulated in vivo and synthetic data<br>3D patch with 160 x 160 x 160<br>Training data size: 30,000 | 3D U-net<br>Loss function: L1 loss |
| Liu and Koch[58] | Simulated in vivo and synthetic data<br>3D patch with 128 x 128 x 64<br>Training data size: 5,000 | 3D U-net<br>Loss function: L1 loss |
| Liu, J. et al.[74] | Synthetic data<br>3D patch with 128 x 128 x 128<br>Training data size: 10,000 | 3D U-net<br>Loss function: L1 loss |
| Liu, Z. et al.[76] | In vivo MEDI<br>3D path with 128 x 128 x 24<br>Training data size: 4,199 | 3D U-net<br>Loss function: L1 loss |
| Polak et al.[61] | In vivo COSMOS<br>3D patch with 176 x 176 x 160<br>Training data size: 40 | 3D Variational network<br>Loss function: L2 loss |
| Jung et al.[56] | In vivo COSMOS and simulated data<br>3D patch with 64 x 64 x 64 | 3D U-net<br>Loss function: L1 loss + gradient different loss |

|  | Training data size: 33,600 | + model loss |
| --- | --- | --- |
| Jochmann et al.[73] | Synthetic data<br>3D patch with 96 x 96 x 96<br>Training data size: 2,100 | 3D U-net<br>Loss function: variance weighted L2 loss |
| Jochmann et al.[83] | Synthetic data<br>3D patch with 96 x 96 x 96<br>Training data size: n/a | 3D U-net<br>Loss function: L2 loss |
| Zhang and Bao[78] | In vivo COSMOS<br>3D patch with 64 x 64 x 16<br>Training data size: Not reported | 3D U-net with spatially-adaptive normalization<br>Loss function: L1 loss + gradient loss + generative adversarial loss |
| Zhang et al.[77] | In vivo COSMOS<br>3D patch with 64 x 64 x 32<br>Training data size: 12,025<br>(Fine-tuned with whole brain) | 3D U-net<br>Loss function: L1 loss + gradient different loss + model loss<br>Fine-tuned with model loss |

**Phase to QSM**

| Geßner and Meineke[84] | Synthetic data<br>3D patch with 128 x 128 x 128<br>Training data size: Not reported | 3D U-net<br>Loss function: Not reported |
| --- | --- | --- |
| Heber et al.[65] | Synthetic data<br>3D patch with 128 x 128 x 128<br>Training data size: 1,000 | Two stacks of u-shaped sub-networks<br>Loss function: L1 loss |
| Liu and Koch[59] | In vivo and simulated background field<br>3D patch<br>Training data size: 5,000 ~ 6,000 | 3D U-net<br>Loss function: L1 loss |
| Kim et al.[66] | In vivo COSMOS and simulated data<br>3D patch with 64 x 64 x 64<br>Training data size: 17,160 | 3D U-net<br>Loss function: L1 loss |
| Wei et al.[68] | In vivo STAR-QSM<br>3D patch with 64 x 64 x 64<br>Training data size: 16,700 | 3D U-net<br>Loss function: L2 loss |

**Figure 1** - The QSM processing pipeline includes phase unwrapping, background field removal (requiring a mask separating background and organ of interest) and the dipole inversion.

**Figure 2** - A comparison of dipole inversion techniques: From left, COSMOS QSM map (gold standard QSM map using multiple orientation data), two different deep learning-based QSM maps (QSMnet$^+$ trained using in-vivo and simulated data, and DeepQSM trained using synthetic data; single orientation field map is used as input for both methods), and three different conventional QSM maps (TKD, iLSQR, and MEDI; single orientation field map is used as input) are shown. Deep learning QSM maps show superior image quality with little streaking artifacts. Computational times for deep learning methods are a few seconds whereas those of the iterative reconstruction methods (iLSQR and MEDI) are several minutes.

**Figure 3** - The effects of the susceptibility range in deep learning QSM. (a) The susceptibility distribution of healthy volunteers used for network training. Since only healthy volunteer data are used for training, the susceptibility values of pathological tissues (e.g. hemorrhage or calcification) are not covered (red arrows). (b) The scatter plot of the susceptibility values in the hemorrhagic lesion (green circle) reconstructed by conventional QSM (x-axis) and deep learning QSM (QSMnet; y-axis). Compared to the conventional QSM, deep learning QSM results underestimated the susceptibility values of the lesion.

**Figure 4 -** The QSM results of the three networks (Unet$_{In-vivo}$, Unet$_{Simul}$, and Unet$_{In-vivo+Simul}$) with different types of training data (in-vivo vs. simulated vs. combined data) are displayed. The simulated field map is generated by the dipole convolution of the COSMOS QSM map. The last column shows the gold standard susceptibility map reconstructed by COSMOS. Normalized root mean squared errors (NRMSE) with respect to the gold standard are displayed in the lower right side of each map. When the type of the input field matches that of the training data, the networks generated lowest NRMSE (green boxes). By using both in-vivo and simulated data for training, the network generalizes better for multiple input types (yellow boxes).

**Figure 5 -** The effects of image resolution on deep learning QSM. The original test data were acquired in 1 mm isotropic resolution and then resampled to three different resolutions (2 mm isotropic, 0.5 mm isotropic, and 0.5 x 0.5 x 1.5 mm$^3$ resolutions). The performance of the network degrades when the resolution of the test data differs from the training data. NRMSE with respect to the COSMOS map is shown at the bottom right corner of each map.

**Figure 6 -** The effects of SNR (a) and B$_0$ direction (b) on deep learning QSM. When the SNR decreases or the B$_0$ direction of test data are different from those of the training data, the performance of the network degrades. NRMSE with respect to the COSMOS map is shown at the bottom left corner of each map.